\documentclass[prd, twocolumn, nofootinbib, floatfix]{revtex4-2} 

\usepackage{amsmath}
\usepackage{graphicx}
\usepackage{dcolumn}
\usepackage{bm}
\usepackage{epsfig}
\usepackage{amssymb,latexsym,mathrsfs,empheq}
\usepackage{graphicx}
\usepackage{color}
\usepackage{xcolor} 
\usepackage{mathtools} 
\usepackage{hyperref} 
\usepackage{cleveref} 
\usepackage{soul}

\hypersetup{
    colorlinks=true,
    linkcolor=red,
    citecolor=blue,
} 

\newcommand{\be}{\begin{equation}}
\newcommand{\ee}{\end{equation}}
\newcommand{\bs}{\begin{split}} 
\newcommand{\bea}{\begin{eqnarray}}
\newcommand{\eea}{\end{eqnarray}} 
 
\newcommand{\alb}{\alpha_B}

\newcommand{\om}{\Omega_m} 
\newcommand{\ode}{\Omega_{\rm de}}

\newcommand{\rde}{\rho_{\rm de}} 
\newcommand{\pde}{P_{\rm de}} 
\newcommand{\rdef}{\rho_{\rm de,eff}}

\newcommand{\gm}{G_{\rm matter}} 
\newcommand{\gl}{G_{\rm light}} 
\newcommand{\geff}{G_{\rm eff}}
 
\newcommand{\mpl}{M_{\rm Pl}^2}

\begin{document}

\title{Cosmology after Phantom Crossing by Horndeski Gravity} 

\author{Eric V.\ Linder} 
\affiliation{
Berkeley Center for Cosmological Physics \& Berkeley Lab, 
University of California, Berkeley, CA 94720, USA
} 

\begin{abstract} 

One possible way to explain the observed effective dark 
energy equation of state crossing $w=-1$ (the phantom 
divide) is through modified gravity. A key point is to 
not view the expansion history in isolation but to take 
into account the other gravitational impacts on growth 
of large scale structure, lensing, etc. Within shift 
symmetric Horndeski gravity this implies three main paths 
for the late time cosmic expansion. All require unusual 
kinetic structure and we analyze their various 
implications for how $w$ should behave after phantom crossing. 

\end{abstract} 

\date{\today} 

\maketitle

\section{Introduction} 

Current cosmic data points toward a ``beautifully 
bizarre'' revolutionary picture of dark energy that 
is not only dynamic but evolves across the phantom 
divide, from effective equation of state ratio 
$w<-1$ at high redshift to $w>-1$ at late 
times \cite{desibao,desilya,deside}. 
Conventional (noninteracting, canonical, general 
relativistic) dark energy cannot achieve 
this \cite{vikman,caldwell,sen05,caldlin}. 

Both interacting and modified gravity models designed 
to enable this behavior of the cosmic expansion tend 
to run afoul of cosmic growth (see \cite{intx} 
and references therein). That is, 
if they couple in some way to matter (either through 
direct interactions, e.g.\ \cite{trodden,chen}, 
or ``phake phantoms'' \cite{scherrer,axion,caldlin}) 
they will significantly alter the growth of large 
scale structure \cite{intx}. On the other hand if they 
couple nonminimally to gravity, modifying general relativity 
(e.g.\ \cite{wolf,intx}), they again impact cosmic growth, 
and possibly also light propagation (lensing). 

Here we consider modified gravity but seek to temper  
the impact on growth and lensing, as well as focusing 
on how the dark energy behaves at late times. That is, 
does the rapid rise of $w$ after crossing $-1$ continue, 
level off, or return to $w=-1$? 

In Section~\ref{sec:expgro} we highlight the 
key aspect of considering cosmic expansion and 
gravitational effects, e.g.\ on growth of large 
scale structure and on lensing, simultaneously. 
Section~\ref{sec:nodom} looks at the implications 
if gravitational braiding effects are large, 
and how dark energy then must cede its dominance. 
For the observed condition of dark energy 
dominance, Section~\ref{sec:nat} explores the 
future behavior of the dark energy equation of 
state under ``natural'' conditions, while 
following sections consider deviations from 
naturalness (Section~\ref{sec:devk}: kinetic 
structure; Section~\ref{sec:devphi}: field 
evolution; Section~\ref{sec:devslow}: field 
coasting; Section~\ref{sec:nok}: vanishing 
kinetic term). We conclude in 
Section~\ref{sec:concl}.

\section{Connecting Cosmic Expansion and Growth} \label{sec:expgro} 

Horndeski gravity provides an excellent framework 
for exploring modified gravity, being highly 
general and well behaved. Observations have constrained 
some of the generality, in particular removing the 
$G_5$ term in the action and restricting the $G_4$ 
term to being a function of the scalar field $\phi$ but 
not its derivative $X\equiv\dot\phi^2/2$ (in the 
most straightforward interpretation of gravitational 
waves propagating at the speed of light). Thus one has 
a Lagrangian
\be 
\mathcal{L}=G_4(\phi)\,R+K(\phi,X)-G_3(\phi,X)\,\Box\phi\ , 
\ee 
where $K$ is the kinetic term. 

Theory motivates restriction to shift symmetry in 
order to rein in quantum corrections. If one adopts 
this then one is left with 
\be 
\mathcal{L}=\frac{1}{2}R+K(X)-G_3(X)\,\Box\phi\ , 
\ee 
normalizing the Planck mass to $\mpl=1$. 
Note that since $G_4(\phi)$ is what led to issues 
with cosmic growth as seen in \cite{wolf,intx}, 
the shift symmetry has an added benefit of 
easing the fit to observations. 

From this action the equations of motion yield the 
modified Friedmann equations 
\bea 
3H^2&=&\rho_m+\rdef\ ,\\  
-2\dot H&=&\rho_m+P_m+\rdef+P_{\rm de,eff}\ ,\\  
0&=&\ddot\phi\,\left[K_X+2XK_{XX}+6H\dot\phi g_X\right]\notag\\ 
&\qquad&\quad+3H\dot\phi K_X+6g\left(\dot H+3H^2\right)\ . \label{eq:fd} 
\eea 
Hereafter we will simply write dark energy rather than 
effective dark energy and suppress the eff subscript. 
We take the matter component to be pressureless, $P_m=0$, 
and it follows the usual continuity equation, avoiding 
a ``phake phantom'' as well as various impacts on large 
scale structure growth. 

The third equation in the set is the scalar field 
equation of motion. Here $g\equiv XG_{3X}$ and a 
subscript $X$ denotes a derivative with respect to $X$. 
We can readily see that the equation  
reduces to the usual Klein-Gordon equation in 
general relativity, i.e.\ $K=X$, $g=0$. 

The explicit forms of the dark energy density and 
pressure are 
\bea 
\rde&=&-K+2XK_X+6H\dot\phi g\ ,\label{eq:rho}\\  
\pde&=&K-2g\ddot\phi\ . \label{eq:pde} 
\eea 
The dark energy equation of state is $w\equiv\pde/\rde$. 
We can see that there is considerable freedom in 
allowing $w$ to cross $-1$ due to the free functions 
$K(X)$, $G_3(X)$ or $g(X)$, and the scalar field 
evolution $\phi(t)$. 

For the modified gravity strengths, i.e.\ the 
effective couplings in the equivalent of the 
Poisson equations for matter perturbations and for 
light propagation, we have 
\be 
\geff=\gm=\gl\ . 
\ee 
That is, such a shift symmetric theory has no 
gravitational slip, so the two metric potentials 
$\Phi$ and $\Psi$ are equal and $\gm=\gl$, 
as in general relativity (GR). However, note 
this does not mean they are equal to the GR 
gravitational strength, Newton's constant (here 
normalized to one). It is 
convenient to write $\geff$ in terms of the 
dimensionless braiding function \cite{bellsaw} 
\be 
\alb=\frac{2\dot\phi g}{H}\ , \label{eq:alb} 
\ee 
leading to 
\be 
\geff=1+\frac{\alb^2}{\alb(2-\alb)+2\alb'}\ , \label{eq:geff} 
\ee 
where a prime denotes $d/d\ln a$. 

Now let us rewrite the the dark energy density and 
pressure to identify key ratios, and highlight 
the close 
interconnection between the modified gravity effects 
and the cosmic expansion. We have 
\bea 
\rde&=&-K\,\left[1-\frac{2XK_X}{K}-\alb\frac{3H^2}{K}\right]\notag\\ 
&=&-K\,\left[1-\frac{2XK_X}{K}-\alb\frac{3H^2}{\rde}\frac{\rde}{K}\right]\ ,\label{eq:rdea}\\ 
\pde&=&K\,\left[1-\frac{2g\ddot\phi}{K}\right]\notag\\ 
&=&K\,\left[1-\alb\frac{\ddot\phi}{3H\dot\phi}\frac{3H^2}{\rde}\frac{\rde}{K}\right]\ . \label{eq:pdea} 
\eea 

We can see that the key ratios are $\rde/(3H^2)=\ode$, 
$\rde/K$, $XK_X/K$, and $\ddot\phi/(3H\dot\phi)$. 
Different cases will arise depending on whether they are 
much smaller, larger, or comparable to unity. 
(Note that in Section~\ref{sec:nok} we show that 
taking $K=0$ will not lead to a viable theory.) 
Connecting the dark energy equation of state, 
in terms of the ratio of its pressure to energy density, 
to the modified gravity braiding $\alb$ is 
important for a synoptic picture of viable cosmology. 
To preserve $\geff\approx1$, i.e.\ no large deviations 
in gravity from GR -- as seen by cosmic growth and lensing 
data -- we require $\alb\ll1$. 
(Also see \cite{nrg}.)

\section{Dark Energy {\it Non-\/}Domination} \label{sec:nodom} 

Let's begin by looking at the $\alb$ term in $\rde$, 
Eq.~(\ref{eq:rdea}). Suppose that it dominates the 
other terms. This gives $\rde=\alb\,(3H^2)$, i.e.\ 
$\ode=\alb$. Since $\alb\ll1$ it is impossible for 
dark energy to dominate. Thus over the main redshift 
range constrained by data, $z\approx[0,1]$, the 
$\alb$ term cannot dominate in $\rde$. 

However we are also interested in the late time 
expansion, to explore how dark energy behaves in 
the future, well after its phantom crossing. If 
$w$ evolves so far as to cross 0 in the future, 
then the dark energy 
density will eventually decline relative to the matter density. 
If the $\alb$ term in $\pde$ also is the dominant 
term then 
\be 
w\to \frac{-\ddot\phi}{3H\dot\phi}\ . \label{eq:wnd1}
\ee 
The field would need to decelerate, i.e.\ 
$\ddot\phi$ and $\dot\phi$ having opposite signs, 
to give $w>0$. 

While if $\pde$ is dominated by the first term, 
$K$, then 
\be 
w\to \frac{K/\rde}{\alb/\ode}\to\frac{K}{\rde}\ . \label{eq:wnd2}
\ee 
Recall from the first paragraph of this section 
that when $\rde$ is dominated by the $\alb$ term 
then $\alb/\ode=1$. Since $\rde$ is 
not dominated by its $K$ term then the equation 
of state must be close to zero in 
this case\footnote{There is 
one exception. Since $\rde$ actually involves 
$-K+2XK_X$ then it is possible for the $\alb$ 
term to be greater than this quantity, but not 
greater than $K$ alone. This requires 
$K\sim X^{1/2}$. It is thus possible that 
$K/\rde$, and hence $w$, could be of order one, 
or even greater than one.\label{ft:kx}}.

\section{Dark Energy Domination Naturally} \label{sec:nat} 

Now we turn to the case of dark energy domination, 
$\rde/(3H^2)\approx1$. We start with the ``natural'' 
case where the other key ratios are also of order one. 
In this case the $\alb$ terms in both $\rde$ and $\pde$ 
are not significant, and 
\be 
w\approx\frac{-1}{1-2XK_X/K}+\mathcal{O}(\alb)\ . \label{eq:wnat} 
\ee 
Defining $n\equiv XK_X/K$, for dark energy to dominate 
we require $w<0$ and hence $n<1/2$. 
(Thus the canonical $K=X$, i.e.\ $n=1$, is not allowed.) 

We then see that $\rde\approx K(2n-1)$ requires $K<0$ 
for positive dark energy density (and negative 
dark energy pressure). This class can also cross 
$w=-1$ when $n$ changes sign, e.g.\ moving from 
$w<-1$ to $w>-1$ as it goes from $n>0$ to $n<0$. 
(Although dark energy doesn't fully dominate when 
data suggest the crossing occurs, $z\approx0.5$, 
one still has $\rde/(3H^2)\sim\mathcal{O}(1)$ then.) 

One might be concerned that 
without $\alb$, and hence $g$, appearing, this seems 
to be simply a k-essence model. Recall that pure 
k-essence alone cannot cross the phantom divide and 
will have a ghost when the kinetic term is negative. 
This does not automatically hold here. The $g$ term 
also enters in the no ghost condition, 
$\alpha_K+(3/2)\alb^2\ge0$, in both $\alb$ and the 
kineticity $\alpha_K$ \cite{bellsaw}. 
Moreover, $g$ and $\alb$ 
also enter in $\geff$, so even in the limit where 
the density and pressure are dominated by the kinetic 
term the theory and its observational implications 
lie outside k-essence.

\section{Deviating the Kinetic Structure} \label{sec:devk} 

Moving away from the natural case where all 
ratios are of order one, 
suppose $XK_X/K$ is not of order one. If it is 
very small, this is of no concern since then simply 
$\rde\approx -K$ and $w\approx-1$. However large 
$XK_X/K$ 
requires special attention. This is because while 
it enters in $\rde$, only $K$ enters in $\pde$. 
If $\rde/K$ becomes large enough, then the $\alb$ 
term in the pressure could dominate over the first 
term. 

Consider a Dirac-Born-Infeld (DBI) kinetic term 
(see e.g.\ \cite{muk,bento}). Then 
\bea 
K(X)&=&-\sqrt{1-X/X_\infty}\\ 
n&\equiv&\frac{XK_X}{K}=\frac{-X/X_\infty}{2(1-X/X_\infty)}\ , \label{eq:ndbi} 
\eea 
and $n\to -\infty$ as $X\to X_\infty$.  As seen by 
Eq.~(\ref{eq:wnat}) this would drive $w\to0^-$, 
or at least to $\mathcal{O}(\alb)$. 
Thus an interesting scenario for the future 
of dark energy is that it could end up scaling 
as matter.

\section{Deviating the Field Evolution} \label{sec:devphi} 

Changing $\ddot\phi/(3H\dot\phi)$ from order one 
also has interesting consequences. Consider 
$\ddot\phi\gg 3H\dot\phi$. If the cosmic expansion 
rate $H$ is the only timescale then this cannot 
be achieved, so such a condition requires 
introduction of a new time scale. 

Recall that in the standard 
Klein-Gordon equation one has the acceleration 
term $\ddot\phi$, the friction term $3H\dot\phi$, 
and the driving term from the steepness of the 
potential, $-dV/d\phi$. 
As clearly illustrated in \cite{paths} for the 
quintessence case, none of these terms dominate over 
the others (unlike in slow roll inflation). 
Here we have no potential, 
but the driving term arises from $-6g(\dot H+3H^2)$. 

One needs an extra ingredient entering the 
driving term, an additional, shorter, time scale 
than the expansion time scale $H^{-1}$. 
One example is a mass scale $M$, e.g.\ 
\be 
g(X)=\gamma(X)+M\ . \label{eq:mass}
\ee 

However this appears problematic as a large 
$\ddot\phi$, increasing at faster than the Hubble 
rate, will eventually lead to a similarly large 
$\dot\phi$ and hence a large $\alb\sim \dot\phi g/H$. 
That will cause large deviations from GR, with 
$\geff$ possibly diverging and instability from 
likely negative sound speed of scalar perturbations, 
$c_s^2<0$ (see \cite{nrg}). 
So this case appears problematic.

\section{Deviating by Slow Roll} \label{sec:devslow} 

We can take the opposite limit of field evolution: 
$\ddot\phi=0$, known as slow roll in inflation. 
Recall that we absolutely 
cannot do this for generic quintessence when the 
dark energy does not fully 
dominate \cite{paths}, i.e.\ as in our present or 
past universe where there is matter. Taking 
$\ddot\phi=0$ would imply $\phi\sim t$, i.e.\ 
the only clock is the field, hence the field must 
dominate the cosmic expansion. 

Suppose the $\ddot\phi=0$ condition holds, at 
some time at least. 
The field equation (\ref{eq:fd}) then has the 
interesting property that it determines the 
action function $g$ in terms of $K$ (though only 
in this limit), 
\be 
g=\frac{-\dot\phi K_X}{3H(1-w_{\rm dom})}\ , \label{eq:gk} 
\ee 
where we have used $\dot H=(-3/2)(1+w_{\rm dom})H^2$, 
and $w_{\rm dom}$ indicates the equation of state 
of the dominant energy density 
component\footnote{This 
solution is 
quite similar to the Branch A solution of 
\cite{2012.03965} that can well temper, 
i.e.\ cancel, a high energy cosmological 
constant, when we have a de Sitter state, with 
$w_{\rm dom}=-1$, $H\to h=\,$const. 
However canceling $\Lambda$ only 
at late times is not so useful.}. 

Keeping $\alb\ll1$ will imply $w\to-1$ when 
dark energy dominates, as 
\be 
\alb\equiv\frac{2\dot\phi g}{H}=\frac{-4XK_X}{3H^2(1-w)}\approx\frac{4n}{(1-2n)(1-w)}\ . 
\ee 
Thus $\alb\ll1$ requires $|n|\ll1$, and hence 
by Eq.~(\ref{eq:wnat}) that $w\to-1$. 
This implies that after dark 
energy crosses the phantom divide to $w>-1$, in this 
specific $\ddot\phi=0$ future it must turn around 
and go back to a de Sitter limit.

\section{No Kinetic Term} \label{sec:nok} 

We know that the $g$ (i.e.\ $G_3$) term must 
exist, otherwise we are left with k-essence, 
which cannot cross the phantom divide by itself. 
Let us consider whether we can do without the 
$K$ term, having only the $G_3$ term in the action. 

When $K=0$, the field equation gains an 
interesting structure, 
\bea 
0&=&\ddot\phi\ 6H\dot\phi g_X+6g\left(\dot H+3H^2\right)\\ 
&=&6a^{-3}\left[(a^3 H)\dot g+g\left(a^3H\right)\dot{\!}\ \right]\ , 
\eea 
since $\dot g=g_X\dot X=g_X\dot\phi\ddot\phi$. 
The solution is 
\be 
g\sim\left(a^3H\right)^{-1}\ . \label{eq:nokga}
\ee 
Once one specifies a $g(X)$ then one can derive $X(a)$. 

However, we see that $\rde=6H\dot\phi g=\alb\,3H^2$, 
and so that dark energy cannot dominate without 
giving a large gravitational deviation. The 
equation of state is $w=-\ddot\phi/(3H\dot\phi)$, 
and must be positive. Thus the combination 
of expansion information (phantom crossing) and 
gravity information (no large deviation from GR) 
requires that both $K$ and $G_3$ terms must enter 
the action.

\begin{table*}[tb]
    \centering
\begin{tabular}{|l|c|c|c|c|c|c|}
\hline 
Case & Eq. & $w$ & Notes \\ 
\hline 
\rule{0pt}{1.1\normalbaselineskip}NonDom1\ & (\ref{eq:wnd1}) & \ $-\ddot\phi/(3H\dot\phi)>0$\ & \ Late time DE no dominate\ \\  
\rule{0pt}{1.1\normalbaselineskip}NonDom2\ & (\ref{eq:wnd2}) & $\approx0^+$ & \ Late time DE no dominate\ \\  
\rule{0pt}{1.1\normalbaselineskip}NonDom3\ & ${}^{\ref{ft:kx}}$ & \ $K/\rde>0$\ & \ Late time DE no dominate; $K\sim X^{1/2}$\ \\  
\rule{0pt}{1.1\normalbaselineskip}{\bf Natural} & (\ref{eq:wnat}) & \ $-1/(1-2n)$\ & \ DE dominate; $n<1/2$; $K<0$\ \\  
\rule{0pt}{1.1\normalbaselineskip}DBI & (\ref{eq:ndbi}) & \ $\approx 0^-$\ & \ DE dominate; $n\to-\infty$\ \\  
\rule{0pt}{1.1\normalbaselineskip}MassAdd\ & (\ref{eq:mass}) & \ ?\ & \ $|\ddot\phi/(3H\dot\phi)|\gg1$; diverging, unstable?\ \\  
\rule{0pt}{1.1\normalbaselineskip}Coast\ & (\ref{eq:gk}) & \ $-1$\ & \ $\ddot\phi=0$; $n\approx0$\ \\  
\rule{0pt}{1.1\normalbaselineskip}No $K$\ & (\ref{eq:nokga}) & \ $-\ddot\phi/(3H\dot\phi)>0$\ & \ $K=0$, $g\sim(a^3H)^{-1}$; unviable\ \\  
\hline 
\end{tabular} \\  
\caption{Different cases enabled by the shift 
symmetric Horndeski gravity. The value of $w$ 
listed is the late time, future value. The 
quantity $n\equiv XK_X/K$. 
The most promising case is the natural case 
of Section~\ref{sec:nat}, enabling both a 
phantom crossing and a diversity of late time 
asymptotes for $w$. 
}
\label{tab:sum} 
\end{table*} 

\section{Conclusions} \label{sec:concl} 

The beautifully bizarre behavior of dark energy 
in rapidly rising in energy density, crossing the 
phantom divide, and evolving to a less negative 
equation of state than $w=-1$ may require a 
beautifully bizarre theory. A major class of 
physics that can enable this behavior is modified 
gravity. 

By combining observational and theoretical 
motivations, leading to consideration of shift 
symmetric, cubic Horndeski gravity, and using the 
key strategy of simultaneously incorporating both 
cosmic expansion and cosmic structure data 
indications -- phantom crossing, dark energy 
domination, and small deviations from GR -- 
we have analyzed the restrictions and outcomes 
of this class of modified gravity. 

We have demonstrated solutions where dark energy 
continues to evolve away from $w=-1$, and indeed 
can have $w>0$ so the dark energy fades away faster 
than matter, where it turns around and restores to 
$w=-1$ at late times, or where dark energy will 
end up at some value $-1<w<0$ (for example, the 
``natural'' case with $K\sim X^{-1}$, i.e.\ $n=-1$, 
goes to $w=-1/3$), or scaling as matter, 
$w\approx0$. The natural case can also provide 
the phantom crossing when $n$ evolves 
from negative to positive. 
Table~\ref{tab:sum} summarizes many of the results. 

Note that the existence of phantom crossing 
and the nonexistence of large deviations from 
GR work together to ensure that both $K$ and 
$G_3$ terms must appear in the action. 
Such shift symmetric, cubic Horndeski theories 
are in the class of No Run Gravity \cite{nrg}, 
and have some interesting properties. To avoid 
Laplace instability in the scalar perturbations, 
the sound speed squared must be nonnegative, 
$c_s^2\ge0$, and this implies $\alb\ge0$ (see the 
Appendix in \cite{nsg}). We see that in many cases 
here $\alb>0$ is also required for positive 
energy density. 

Note that $\alb\ge0$ has the implication that 
$\geff\ge1$, as seen from Eq.~(\ref{eq:geff}) 
and \cite{nrg}; indeed, using the stability 
condition that \cite{nrg} derived, 
\be 
0\le\geff(a)-1\le \frac{\alb(a)}{3\om(a)}\ . 
\ee 
Thus, in the range best covered by observations, 
and in particular around the phantom crossing 
time, $\geff-1\ll1$ so $\alb\ll1$ is required. 

It is interesting that this class of theories 
strengthens gravity, giving a clear prediction 
for large scale structure surveys. Together 
with its requirement of no gravitational slip, 
i.e.\ the behaviors of clustering and of lensing  
are closely tied together, $\gm=\gl$, this offers 
clear tests for shift symmetric, cubic 
Horndeski gravity as the physics behind the 
phantom crossing.

\acknowledgments 

I thank Bob Scherrer for coining the term 
``phake phantoms''.

\end{document}